\documentclass[12pt]{article}

\usepackage[margin=1in,nohead,dvips,a4paper]{geometry}
\usepackage{times,pstricks,pst-node}

\psset{linewidth=.5pt,dash=4pt 4pt,unit=.15in,arrowsize=2pt 3}

\def\arrowLine(#1,#2)(#3,#4){%
  \pcline(#1,#2)(#3,#4)%
  \lput{:U}{
    \pspicture(0,0)(0,0)
      \psline[arrows=->](2.3pt,0)(2.4pt,0)
    \endpspicture
  }
}
\begin{document}

\title{1/N phenomenon for some symmetry classes of the odd alternating sign matrices}
\author{Yu.~G.~Stroganov\\
\small \it Institute for High Energy Physics\\[-.5em]
\small \it 142281 Protvino, Moscow region, Russia}
\date{}

\maketitle

\begin{abstract} We consider the alternating sign matrices of the odd order 
that have some kind of central symmetry. Namely, we deal with matrices invariant under 
the half-turn, quarter-turn and flips in both diagonals.
In all these cases, there are two natural structures in the centre of the matrix.
For example, for the matrices invariant under the half-turn the central element 
is equal
$\pm 1$. It was recently found that $A^+_{HT}(2m+1)/A^-_{HT}(2m+1)$=(m+1)/m.
We conjecture that similar very simple relations are valid in the two remaining cases. 
\end{abstract}

\vskip 1em

An alternating-sign matrix is a matrix with entries $1$, $0$, and $-1$ such
that the $1$ and $-1$ entries alternate in each column and each row and
such that the first and last nonzero entries in each row and column are
$1$. Starting  from the famous conjectures by Mills, Robbins and Rumsey
\cite{MilRobRum82, MilRobRum83} and \cite{Rob00} a lot of enumeration and 
equinumeration
results on alternating-sign matrices and their various subclasses were
obtained. Most of the results were proved using bijection between matrices
and states of different variants of the statistical square-ice model. For
the first time such a method to solve enumeration problems was used by
Kuperberg~\cite{Kup96}, see also the rich in results paper~\cite{Kup02}.

In the recent paper~\cite{RazStr05} that was mentioned 
in the abstract, authors studied enumerations of the half-turn
symmetric alternating-sign matrices of odd order on the base of the
corresponding square-ice model. 

We say that an alternating-sign matrix $ A$ of the order $n$ is half-turn
symmetric if 
\begin{eqnarray}
\label{defHT}
&& (A)_{n+1-i,n+1-j}=(A)_{i,j}, \quad  i,j=1,...,n. \nonumber
\end{eqnarray}

It appears that one can separate the contributions to
the partition function of the states corresponding to the alternating-sign
matrices having~$1$ and~$-1$ as the central entry (Theorem 2 of paper~\cite{RazStr05}).
The authors have proved amazingly simple relation 
\begin{eqnarray}
\label{relHT}
&& A^{(+1)}_{HT}(2m+1)/A^{(-1)}_{HT}(2m+1)=(m+1)/m.
\end{eqnarray}

In their next paper~\cite{RazStr06} authors treated the quarter-turn
symmetric alter\-na\-ting-sign matrices of odd order and proved 
the conjectures by  Robbins~\cite{Rob00} related to enumeration of these matrices. 

An alternating-sign  $n \times n$ matrix $ A$ is said to be quarter-turn
symmetric if 
\begin{eqnarray}
\label{defQT}
&& (A)_{j,n+1-i}=(A)_{i,j}, \quad  i,j=1,...,n \nonumber.
\end{eqnarray}
We consider the matrices of odd order and write $n=2m+1$ .
Let $k^{\pm}$ be the numbers 
of the noncentral matrix elements that are equal $\pm 1$ respectively.
It is evident that 
\begin{eqnarray}
\label{central}
&& k^{+}+k^{-}+(A)_{m+1,m+1}=n=2m+1 \nonumber.
\end{eqnarray}
The numbers $ k^{\pm}$ are divisible by 4. One obtains 
that the quarter-turn symmetric alternating-sign matrix of the order $n=2m+1$
has $(A)_{m+1,m+1}=1$ or $(A)_{m+1,m+1}=-1$ as the central entry according to
whether  $m$ is even or odd. 

In this note we report the additional observation related to the ratio 
of numbers of these matrices with the different structures in the centre.

In the case $m=2\mu, \quad n=4\mu+1$, one has $1$ as the central entry.
There are two different structures in the centre: four adjacent entries can be 
$0$ or $-1$. Using evident notations we can describe our observation as follows:

\bf Conjecture 1a \rm
\begin{eqnarray}
\label{relQT1}
&& A^{(0)}_{QT}(4\mu+1)/A^{(-1)}_{QT}(4\mu+1)=(\mu+1)/\mu.
\end{eqnarray}

In the case $m=2\mu+1, \quad n=4\mu+3$, one has $-1$ as the central entry.
Once again, there are two different structures in the centre: four adjacent 
entries can be 
$1$ or $0$. We find now

\bf Conjecture 1b \rm
\begin{eqnarray}
\label{relQT3}
&& A^{(1)}_{QT}(4\mu+3)/A^{(0)}_{QT}(4\mu+3)=(\mu+1)/\mu.
\end{eqnarray}

Let us consider the last case - alternating-sign matrices of odd order 
invariant under flips in both diagonals. Robbins found a pattern for this
case~\cite{Rob00} but the author does not know the corresponding proof.
As in the case of the half-turn symmetric alternating-sign matrices,
we have ~$1$ or~$-1$ as the central entry and observe
amazingly simple relation 

\bf Conjecture 2 \rm
\begin{eqnarray}
\label{relDD}
&& A^{(+1)}_{DD}(2m+1)/A^{(-1)}_{DD}(2m+1)=(m+1)/m.
\end{eqnarray}
(Compare with equation (\ref{relHT}).)

{\it Acknowledgments}
The author warmly thanks Alexander Razumov for numerous useful
discussions.
The work was supported in part by the Russian
Foundation for Basic Research under grant \# 07--01--00234.

\end{document}